\documentclass{pasj00}
\draft

\begin{document}
\SetRunningHead{Fujita and Goto}{Evolution of Galaxies in and around
Clusters} 
\Received{2004/04/12}
\Accepted{2004/06/25}

\title{The Evolution of Galaxies in and around Clusters at
High-Redshift}

\author{Yutaka \textsc{Fujita}} 
\affil{National Astronomical
Observatory, \\and \\Department of Astronomical Science, The Graduate
University for Advanced Studies,\\ 2-21-1 Osawa, Mitaka, Tokyo 181-8588}
\email{yfujita@th.nao.ac.jp} 

\and 

\author{Tomotsugu \textsc{Goto}}
\affil{Department of Physics and Astronomy, The Johns Hopkins
University, \\ 3400 North Charles Street, Baltimore, MD 21218-2686, USA}

%


\maketitle

\begin{abstract}
 In this paper, we focus on ram-pressure stripping and evaporation of
 disk galaxies in and around a cluster. We show that the evolution of
 the disk surface density affects the efficiency of ram-pressure
 stripping of galaxies at $z\gtrsim 1$. We also consider the saturation
 of thermal conduction in detail and show that it cannot be ignored at
 larger radii of a cluster, which makes the time-scale of the
 evaporation larger.  Both the ram-pressure stripping and evaporation
 could affect the evolution of galaxies even around a cluster. In
 particular, the observed gradual decline of the star-formation rates of
 galaxies in and around clusters could be explained by evaporation
 without resorting to speculative strangulation (stripping of warm gas
 in galactic halos).
\end{abstract}

\section{Introduction}

Clusters of galaxies in the redshift range of 0.2--0.5 often exhibit an
overabundance, relative to present-day clusters, of blue galaxies
(Butcher, Oemler 1978, 1984). This star-formation activity is often
called the Butcher--Oemler effect. Subsequent studies have confirmed
this trend \citep{cou87,rak95,lub96,mar00,ell01,got03a}.

On the other hand, a correlation has been known between galaxy
morphology and the local environment. \citet{dre80} studied 55 nearby
galaxy clusters and found that fractions of early-type galaxies increase
and those of late-type galaxies decrease with increasing local galaxy
density in the clusters.  Subsequent studies have confirmed the
morphology--density relation, that is, early-type galaxies are dominated
in inner region of clusters where the density is high and the fraction
decreases toward the outside of the clusters \citep{whi91,whi93}.

Recently, we have gradually understood that the above phenomena are
related to each other. \citet{dre94} and \citet{cou98} found that most
of the blue galaxies observed as the Butcher--Oemler effect are normal
spirals with active star formation. \citet{dre97} studied 10 clusters at
$z\sim 0.5$, and found the morphology--density relation at these
redshifts. However, they also found that S0 fractions are much smaller
than those in nearby clusters. The low fractions of S0 galaxies have
also been observed by others \citep{fas00}. In many clusters, their
galaxy population gradually changes from a red, evolved, early-type
population in the inner part of the clusters to a progressively blue,
later-type population in the extensive outer envelope of the clusters
\citep{abr96,bal97,rak97,oem97,sma98,cou98,van98,got04}. These
observations suggest that the blue, normal spirals observed in
high-redshift clusters were originally field galaxies; they fell into
clusters and evolved into the non-blue S0 galaxies observed in nearby
clusters.

Several mechanisms have been proposed that can lead to color and
morphological transformations between galaxy classes in clusters, such
as, galaxy mergers \citep{too72}, tides from the cluster potential
\citep{byr90,fuj98}, tidal interactions between galaxies \citep{moo96},
ram-pressure stripping
\citep{gun72,tak84,gae87,por93,bal94,fuj99a,fuj99b,aba99,mor00,fuj01a},
evaporation \citep{cow77b}, and a gradual decline in the star-formation
rate of a galaxy owing to the stripping of halo gas (strangulation or
suffocation; Larson et al. 1980; Kodama et al. 2001; Bekki et al. 2002).

In this paper, we focus on ram-pressure stripping and evaporation
following Fujita (2004, hereafter Paper~I).  We mostly consider the
star-formation history of galaxies, and do not treat the morphological
transition in detail. We consider the redshift evolution of the
disk-size and the surface density, which were not considered in Paper I.
Moreover, we consider the evaporation in detail, while paying attention
to the saturation of thermal conduction. We mainly treat the
environmental effects on galaxies during the first infall into a
cluster; we do not consider their long-term evolution. Thus, the
high-redshift galaxies we investigate may not be direct progenitors of
galaxies at $z\sim 0$. Although we often use the word `evolution' from
now on, it does not mean the evolution of `a particular
galaxy'. Instead, we discuss the differences of the average properties
of galaxies at low and high redshifts. Although the models presented in
this paper could be included in complicated semi-analytic models of
galaxy formation, it would be instructive to study the characteristics
of the environmental effects using simple models before we consider such
a semi-analytic approach.

This paper is organized as follows. In section~\ref{sec:model}, we
summarize our models. In section~\ref{sec:result} we give the results of
our calculations, and compare them with observations in
section~\ref{sec:disc}. Conclusions are given in section~\ref{sec:conc}.
As a cosmological model, we consider a cold dark-matter model with a
non-zero cosmological constant ($\Lambda$CDM model). The cosmological
parameters are $h=0.7$, where the Hubble constant is given by $H_0=100
h\rm\; km\; s^{-1}\; Mpc^{-1}$, $\Omega_0=0.25$, $\lambda_0=0.75$, and
$\sigma_8=0.8$.

\section{Models}
\label{sec:model}

\subsection{The Growth of Clusters}
\label{sec:growth}

The typical mass of progenitors of a cluster can be derived from the
extended Press--Schechter model (EPS) \citep{bow91,bon91,lac93} and its
further extension \citep{fuj02b}. The latter is a Press--Schechter model
including the effect of spatial correlations among initial density
fluctuations (SPS). The models are summarized in Paper~I.

Cluster progenitors can be classified into two categories: one is the
main cluster and the others are subclusters. The main cluster is the
progenitor that was located near to the center of the current cluster,
and had a much larger mass than the other progenitors. Subclusters are
progenitors other than the main cluster that were located in the
vicinity of the main cluster. Since the SPS model includes information
about the spatial correlation of the initial density fluctuations of the
universe, we can separately derive the typical masses of the main
cluster and the subclusters.  Moreover, the SPS model includes the EPS
model. As shown in Paper~I, we can define the typical mass of the main
cluster at redshift $z$, which becomes a cluster of mass $M_0$ at a
later time, $z_0 (<z)$, as
\begin{equation}
 \label{eq:m_ave_eps}
 \bar{M}_{\rm EPS}(z|M_0,z_0)
=\frac{\int_{M_{\rm min}}^{M_0} M P_{\rm EPS}(M,z|M_0,z_0)d M}
{\int_{M_{\rm min}}^{M_0} P_{\rm EPS}(M,z|M_0,z_0)d M} \:,
\end{equation}
where $M_{\rm min}$ is the lower cutoff mass, and $P_{\rm
EPS}(M,z|M_0,z_0)$ is the conditional probability based on the EPS model
that a particle that resides in an object (`halo') of mass $M_0$ at
redshift $z_0$ is contained in a smaller halo of mass $M\sim M_1+d M$ at
redshift $z$ ($z>z_0$). We choose $M_{\rm min}=10^8\: \MO$, which
corresponds to the mass of dwarf galaxies. The definition of $P_{\rm
EPS}$ is shown in equation~(8) in Paper~I.

The typical mass of the subclusters is given by
\begin{equation}
 \label{eq:m_ave}
 \bar{M}_{\rm SPS}(z|M_0,z_0)
=\frac{\int_{M_{\rm min}}^{M_0} M P_{\rm SPS}(M,z|M_0,z_0)d M}
{\int_{M_{\rm min}}^{M_0} P_{\rm SPS}(M,z|M_0,z_0)d M} \:,
\end{equation}
where $P_{\rm SPS}(M,z|M_0,z_0)$ is the conditional probability based on
the SPS model. The definition of $P_{\rm SPS}$ is shown in equation~(7)
in Paper~I.  We define the radius of the region that later becomes a
cluster of mass $M_0$ at $z=z_0$ as $R_0$. Following Paper~I, we
consider the subclusters that were initially located at $0.7 R_0 < r <
R_0$ in the precluster region. We refer to the inner radius as $R_{\rm
in}$.

\subsection{Ram-Pressure Stripping}
\label{sec:model_ram}

We adopt the ram-pressure stripping model of Paper~I.  In the following
sections, we often refer to a relatively large dark halo containing
galaxies and gas as a `cluster'. This `cluster' includes the main
cluster, the subclusters, and so on.

We assume that a cluster is spherically symmetric and the density
distribution of the dark matter is
\begin{equation}
 \label{eq:rho_m}
\rho_{\rm m}(r)=\rho_{\rm mv}(r/r_{\rm vir})^{-a}\:,
\end{equation}
where $\rho_{\rm mv}$ and $a$ are constants, $r_{\rm vir}$ is the virial
radius of the cluster, and $r$ is the distance from the cluster
center. We choose $a=2.4$ and determine $\rho_{\rm mv}$ and $r_{\rm
vir}$ by a spherical collapse model following Paper~I. We note that the
average mass density of a cluster increases toward high redshift. For
example, it is proportional to $(1+z)^3$ for the Einstein--de Sitter
universe.

We ignore the self-gravity of the ICM, and consider two ICM mass
distributions. When the ICM is not heated by anything other than the
gravity of the cluster, the distribution is written as
\begin{equation}
\label{eq:ICM_G}
 \rho_{\rm ICM}(r)=\rho_{\rm ICM, vir}
\frac{[1+(r/r_{\rm c})^2]^{-a/2}}
{[1+(r_{\rm vir}/r_{\rm c})^2]^{-a/2}}\:,
\end{equation}
where $r_{\rm c}/r_{\rm vir}=0.1$. We call this model `the non-heated
ICM model'. In this model, we assume that the ICM temperature equals to
the virial temperature ($T_{\rm ICM}=T_{\rm vir}$).

However, at least for nearby clusters and groups, an entropy excess of
the ICM (an entropy floor) has been observed in X-rays in the central
regions. This indicates that the ICM has been heated by some sources,
such as AGNs or supernovae, in addition to the gravity of the clusters
and groups \citep{pon99,llo00,mul00,mul03}. Thus, we construct `the
heated ICM model'. We assume that the ICM is heated before cluster
formation. Although there is a debate about whether the heating took
place before or after cluster formation
\citep{fuj01b,yam01,bab02,voi02}, the following results would not be
much different, even if the ICM is heated after cluster formation
\citep{loe00}. If the ICM is heated non-gravitationally before cluster
formation, the final distribution depends on the virial temperature of
the cluster, $T_{\rm vir}$. If $T_{\rm vir}\ge T_0$, we assume that a
shock forms near the virial radius of the cluster. In this study, we
assume $T_0=0.8$~keV from X-ray observations (Fujita, Takahara 2000;
Paper~I). The ICM distribution is given by
\begin{equation}
\label{eq:ICM_H}
 \rho_{\rm ICM}(r)=\rho_{\rm ICM, vir}
\frac{[1+(r/r_{\rm c})^2]^{-3b/2}}
{[1+(r_{\rm vir}/r_{\rm c})^2]^{-3b/2}}\:, 
\end{equation}
where $b=(2.4/3)T_{\rm vir}/T_{\rm ICM}$ and $T_{\rm ICM}$ is the
temperature of the ICM, which is given by $T_{\rm ICM}=T_{\rm
vir}+T_0$. If $T_{\rm vir}< T_0$~keV, a shock does not form, but the gas
accreted by a cluster adiabatically falls into the cluster.  The ICM
density and temperature profiles are approximately given by
\begin{equation}
\label{eq:rho_ad}
 \rho_{\rm ICM}(r)=\rho_{\rm ICM, vir}
\left[1+\frac{3}{A}
\ln\left(\frac{r_{\rm vir}}{r}\right)\right]^{3/2}\:,
\end{equation}
\begin{equation}
\label{eq:T_ad} 
 T_{\rm ICM}(r) = \frac{4}{15}\left[1+\frac{3}{A}
\ln \left(\frac{r_{\rm vir}}{r}\right)\right]\:,
\end{equation}
where $A$ is the constant determined by the adiabat of the ICM (Balogh
et al. 1999a; Paper~I). Since equations~(\ref{eq:rho_ad})
and~(\ref{eq:T_ad}) diverge at $r=0$, we take their values at $r=0.1
r_{\rm vir}$ as the central values. In equations~(\ref{eq:ICM_G}),
(\ref{eq:ICM_H}), and (\ref{eq:rho_ad}), the normalizations of the ICM
profile are given by the observed ICM fraction of clusters or the rate
of gas accretion to clusters (Paper~I). If the non-gravitational heating
makes the accretion time larger than the lifetime of a cluster, the
cluster cannot accrete much gas and the gas fraction is smaller than the
average in the universe. This happens at high redshifts ($z\gtrsim 1$--2)
in our models.

We consider a radially infalling disk galaxy from the turnaround radius
of a cluster ($2r_{\rm vir}$). As the velocity of the galaxy increases,
the ram-pressure from the ICM also increases. The condition of
ram-pressure stripping is
\begin{eqnarray}
  \rho_{\rm ICM}v_{\rm rel}^2 
 & >& 2\pi G \Sigma_{\star} \Sigma_{\rm HI} \nonumber\\
 & =& v_{\rm rot}^2 r_{\rm gal}^{-1} 
  \Sigma_{\rm HI} \label{eq:grav2} \nonumber\\
 & =& 2.1\times 10^{-11}{\rm dyn\: cm^{-2}}
               \left(\frac{v_{\rm rot}}{220\rm\; km\: s^{-1}}\right)^2
               \nonumber\\
 &  &   \times \left(\frac{r_{\rm gal}}{10\rm\; kpc}\right)^{-1}
               \left(\frac{\Sigma_{\rm HI}}
               {8\times 10^{20} 
                   m_{\rm H}\;\rm cm^ {-2}}\right) \label{eq:strip}\:, 
\end{eqnarray}
where $v_{\rm rel}$ is the relative velocity between the galaxy and the
ICM, $\Sigma_{\star}$ is the gravitational surface mass density,
$\Sigma_{\rm HI}$ is the surface density of the H\,{\footnotesize I}
gas, $v_{\rm rot}$ is the rotation velocity, and $r_{\rm gal}$ is the
characteristic radius of the galaxy \citep{gun72,fuj99a}. We define the
cluster radius at which condition (\ref{eq:strip}) is satisfied for the
first time as the stripping radius, $r_{\rm st}$. Since we assume that
the ICM is nearly in pressure equilibrium for $r<r_{\rm vir}$, the
relative velocity, $v_{\rm rel}$, is equivalent to the velocity of the
galaxy relative to the cluster, $v$, for $r<r_{\rm vir}$.

\subsection{Evaporation}

The time-scale of the evaporation of cold gas in a galaxy is
written as
\begin{equation}
\label{eq:cond}
 t_{\rm cond}\approx \frac{3}{2}
\frac{k_{\rm B} T_{\rm ICM}}{\mu m_{\rm H}}
\frac{M_{\rm cold}}{|L|} \:,
\end{equation}
where $k_{\rm B}$ is the Boltzmann constant, $T_{\rm ICM}$ is the ICM
temperature, $\mu$ (=0.6) is the mean molecular weight, $m_{\rm H}$ is
the hydrogen mass, $M_{\rm cold}$ is the mass of cold gas in the galaxy,
and $L$ is the energy flux from the hot ICM surrounding the galaxy via
thermal conduction (Paper~I). We define neutral and molecular gas
confined in a galactic disk as cold gas; we do not consider the cold gas
in a galactic bulge. If the electron mean-free path is smaller than the
spatial scale of the temperature gradient around the galaxy, the thermal
conduction is not saturated and the energy flux is given by
\begin{equation}
\label{eq:Lcond}
 |L_{\rm nsat}| = 4\pi r_{\rm gal}^2 \kappa_0 T_{\rm ICM}^{7/2}\:,
\end{equation}
where $r_{\rm gal}$ is the galaxy radius, and $\kappa_0 = 5\times
10^{-7}\rm\; erg\; cm^{-1}\; s^{-1}\; K^{-3.5}$ (Paper~I). If the
electron mean-free path is larger than the spatial scale of the
temperature gradient, the thermal conduction is saturated and the energy
flux is given by
\begin{equation}
\label{eq:Lsat}
 |L_{\rm sat}| 
= 4\pi r_{\rm gal}^2\times 0.4 n_{\rm e} k_{\rm B} T_{\rm ICM} 
\left(\frac{2 k_{\rm B}T_{\rm ICM}}{\pi m_{\rm e}}\right)^{1/2}\;,
\end{equation}
where $n_{\rm e}$ is the electron number density, and $m_{\rm e}$ is the
electron mass \citep{cow77a}. For convenience, we often refer to $t_{\rm
cond}$ given by $L=L_{\rm nsat}$ as $t_{\rm cond, nsat}$ and that given
by $L=L_{\rm sat}$ as $t_{\rm cond, sat}$. An actual energy flux is
given by $|L|=\min(|L_{\rm nsat}|, |L_{\rm sat}|)$. 

\subsection{Evolution of Disk Properties}

The most important difference between this study and Paper~I is that we
consider the redshift evolution of disk properties in this paper. We
adopted a simple model, discussed in \citet{mo98}, for the galactic
disk.

For a given rotation velocity, the galaxy radius at redshift $z$ is
given by
\begin{equation}
\label{eq:evo_rad}
 r_{\rm gal}(z) = r_{\rm gal, 0}(H[z]/H_0)^{-1}\:,
\end{equation}
where $r_{\rm gal,0}$ is the galaxy radius at $z=0$ and 
\begin{equation}
 H(z) = H_0 [\lambda_0 
+ (1-\lambda_0-\Omega_0)(1+z)^2+\Omega_0(1+z)^3]^{1/2}
\end{equation}
is the Hubble constant at redshift $z$. The surface density and mass
of the disk at redshift $z$ are given by
\begin{equation}
\label{eq:evo_star}
 \Sigma_{\star}(z) = \Sigma_{\star, 0} H(z)/H_0\:,
\end{equation}
\begin{equation}
\label{eq:evo_Md}
 M_{\rm disk}(z) = M_{\rm disk, 0} (H[z]/H_0)^{-1}\:,
\end{equation}
where $\Sigma_{\star, 0}$ and $M_{\rm disk, 0}$ are the surface density
and mass at $z=0$, respectively. As shown in figure~1 in \citet{mo98},
$H(z=1)\sim 2 H_0$. Thus, the disk radius (surface density) at $z\sim 1$
is a factor of two smaller (larger) than that at $z=0$.

We give the column density of the H\,{\footnotesize I} gas in the disk
and the mass of the cold gas {\it before} the galaxy is affected by
environmental effects as follows.  We assume that $\Sigma_{\rm
HI}\propto \Sigma_\star$ for simplicity; in the future study, we will
not assume this by using a semi-analytic model of galaxy
formation. Thus, the column density of the H\,{\footnotesize I} gas at
redshift $z$ is given by
\begin{equation}
\label{eq:evo_HI}
 \Sigma_{\rm HI}(z) = \Sigma_{\rm HI, 0} H(z)/H_0\:,
\end{equation}
where $\Sigma_{\rm HI, 0}$ is the column density at $z=0$. We also
assume that $M_{\rm cold}\propto M_{\rm disk}$ and 
\begin{equation}
\label{eq:evo_Mc}
 M_{\rm cold}(z) = M_{\rm cold, 0} (H[z]/H_0)^{-1}\:,
\end{equation}
where $M_{\rm cold, 0}$ is the mass of the cold gas at $z=0$.

\section{Results}
\label{sec:result}

\begin{figure}
  \begin{center}
    \FigureFile(50mm,50mm){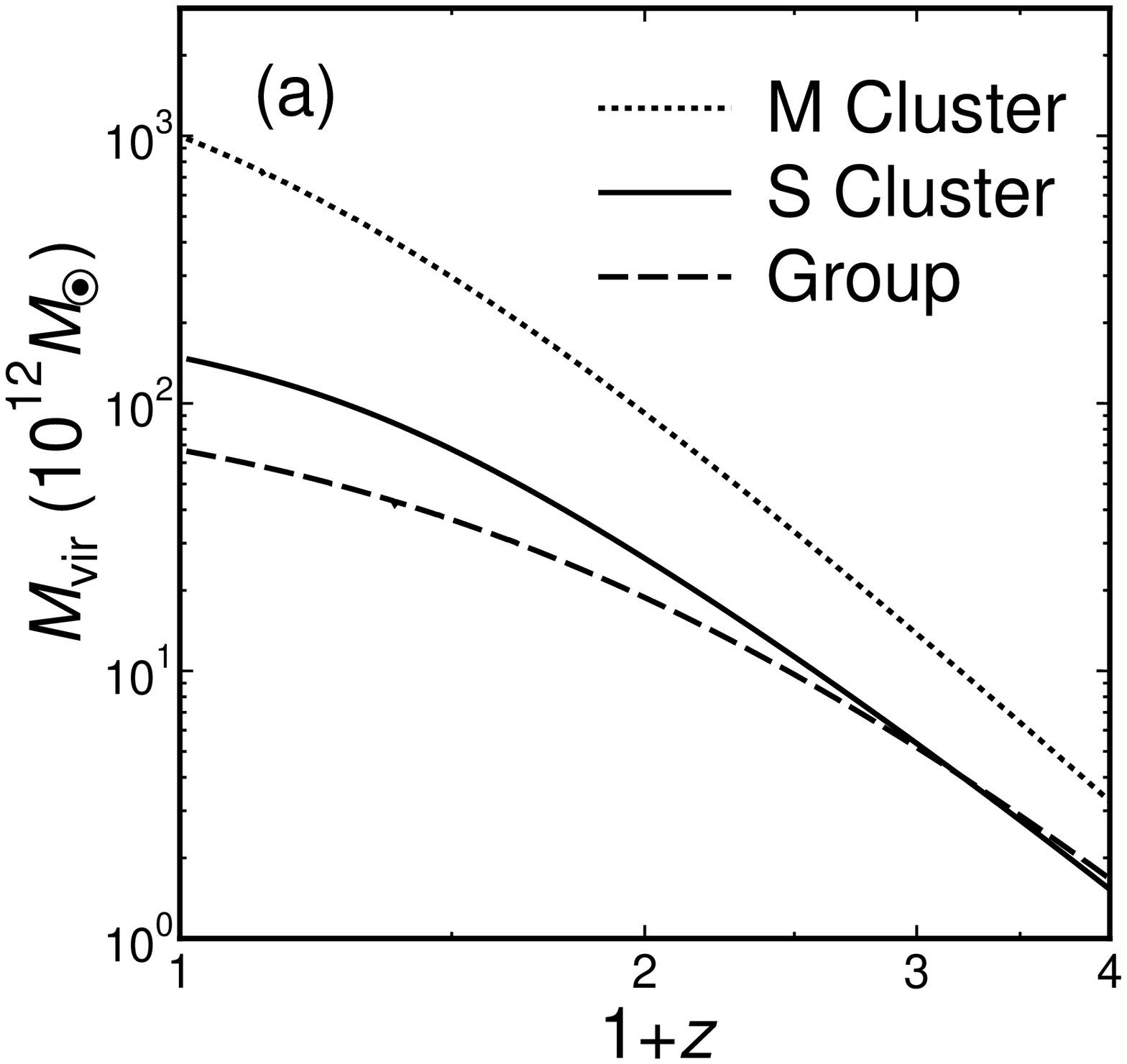}
    \FigureFile(50mm,50mm){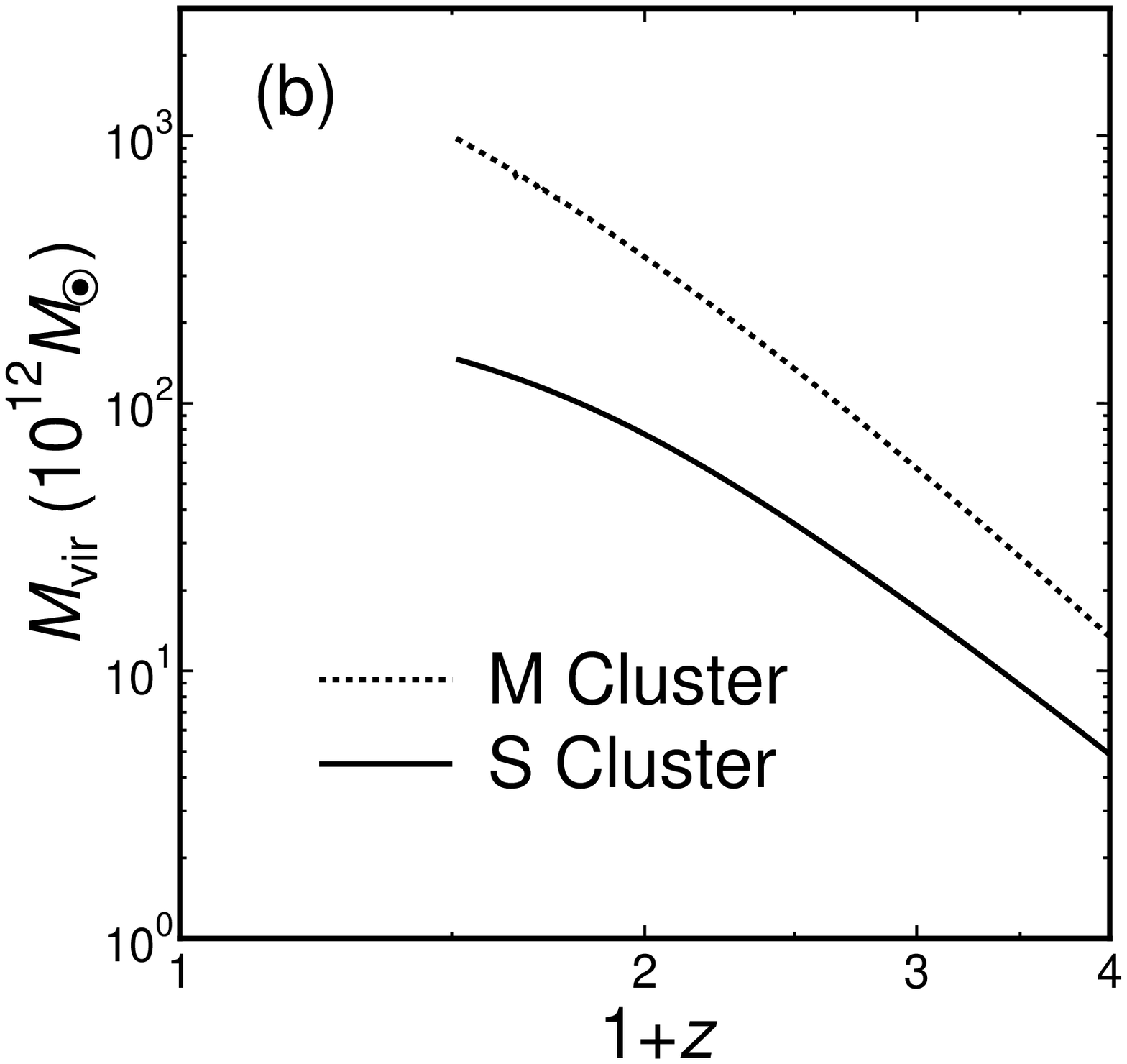}
  \end{center}
  \caption{(a) Mass evolution of the main cluster (dotted line), the
typical subcluster (solid line) of the LCL, and the group (dashed
line). (b) Mass evolution of the main cluster (dotted line) and the
typical subcluster (solid line) of the HCL.} \label{fig:mass}
\end{figure}

We considered the evolution of clusters with three masses at $z=z_0$.
Two of them are the same as those in Paper~I. The masses of the two
clusters at $z_0=0$ are $M_0=1\times 10^{15}\; \MO$ and $6.7\times
10^{13}\; \MO$. We call the former cluster `the low-redshift cluster
(LCL)', which is studied to be compared with clusters observed at $z\sim
0.5$. Actually, at $z=0.5$, the mass of the main cluster is $M_{\rm
vir}=3\times 10^{14}\; \MO$, which is close to the masses of well-known
clusters observed at $z\sim 0.5$ \citep{sch99}. The typical mass of the
subclusters of the LCL at $z=0.5$ is $M_{\rm vir}=6.7\times 10^{13}\;
\MO$ (figure~\ref{fig:mass}). The latter cluster with $M_0=6.7\times
10^{13}\; \MO$ at $z_0=0$ is investigated to be compared with the
subclusters. Since the mass scale of this cluster is that of groups of
galaxies, we call this cluster `the group'. For the group, we considered
only the evolution of the main cluster. The third cluster that we
studied has a mass of $M_0=1\times 10^{15}\; \MO$ at $z_0=0.5$. This
cluster was studied to be compared with recent observations of clusters
observed at $0.5\lesssim z \lesssim 1$. We call this cluster `the
high-redshift cluster (HCL)'.

In figure~\ref{fig:mass}, we present the evolutions of the cluster
masses. We expect that when the mass of the main cluster satisfies the
relation $M_{\rm vir}/M_0 = (R_{\rm in}/R_0)^3$, the subclusters begin
to be included in the main cluster and become subhalos. For the
parameters we adopted (e.g., $R_{\rm in}=0.7 R_0$), the subclusters of
the LCL and HCL are absorbed by the main clusters, together with the
galaxies in them, at $z\lesssim 0.4$ and 1, respectively. This means
that at $z\sim 0.4$ (1) the subclusters of the LCL (HCL) are to be
observed just outside the main cluster of the LCL (HCL).

Since we are interested in galaxies at high redshift, we investigated
relatively large galaxies that can be observed in detail. We consider
two model galaxies following Paper~I. While we fixed the rotation
velocities of the galaxies $v_{\rm rot}$, we changed the radius, surface
density, and H\,{\footnotesize I} column density according to equations
(\ref{eq:evo_rad}), (\ref{eq:evo_star}), and (\ref{eq:evo_HI}). The
parameters for the bigger model galaxy, which is similar to the Milky
Way, are $v_{\rm rot}=220\:\rm km\: s^{-1}$, $r_{\rm gal, 0}=10$ kpc,
and $\Sigma_{\rm HI, 0}=8\times 10^{20}m_{\rm H}\rm\: cm^{-2}$. The
parameters for the smaller model galaxy, which is similar to M33, are
$v_{\rm rot}=105\:\rm km\: s^{-1}$, $r_{\rm gal,0}=5$ kpc, and
$\Sigma_{\rm HI,0}=14\times 10^{20}m_{\rm H}\rm\: cm^{-2}$. These are
the values at $z=0$ even for the galaxies in the HCL. From now on, this
disk evolution is considered unless otherwise mentioned.

\subsection{Ram-Pressure Stripping}
\label{sec:ramp}

\begin{figure}
  \begin{center}
    \FigureFile(100mm,100mm){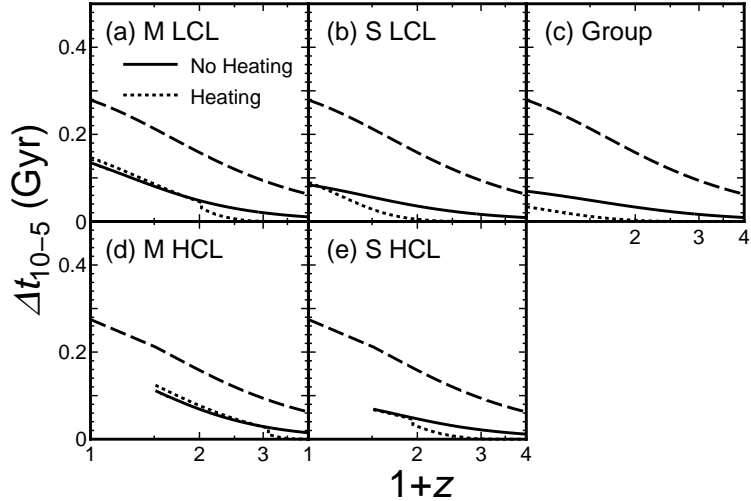}
  \end{center}
  \caption{Time-scale of ram-pressure stripping, $\Delta t_{\rm 10-5}$
for the bigger galaxy in the (a) main cluster of the LCL, (b) subcluster
of the LCL, (c) group, (d) main cluster of the HCL, and (e) subcluster
of the HCL. The solid lines are the results of the non-heated ICM model
and the dotted lines are those of the heated ICM model. The dashed-line
shows the rotation time of the galaxy ($t_{\rm rot}$). \label{fig:delt}}
\end{figure}

First, we consider the time-scale of ram-pressure stripping. Although
equation~(\ref{eq:strip}) is the condition at the representative radius
of a galaxy, numerical simulations performed by Abadi, Moore, and Bower
(1999) showed that it can also be applied at `each radius' of the
galaxy. Assuming that $v_{\rm rot}$ is the constant and $\Sigma_{\rm
HI}\propto \Sigma_\star$, the H\,{\footnotesize I} column density has
the relation $\Sigma_{\rm HI}\propto \Sigma_\star \propto v_{\rm
rot}^2/\tilde{r}\propto \tilde{r}^{-1}$, where $\tilde{r}$ is the
distance from the galaxy center. Thus, equation~(\ref{eq:strip}) shows
that the ram-pressure required for stripping at $\tilde{r}=5$~kpc is
4-times larger than that at $\tilde{r}=10$~kpc. Figure~\ref{fig:delt}
shows the time elapsed since the cold gas at $\tilde{r}=10$~kpc is
stripped until that at $\tilde{r}=5$~kpc is stripped, $\Delta t_{\rm
10-5}$, for the radially infalling bigger galaxy ($v_{\rm rot}=220\rm\:
km\: s^{-1}$). For $\tilde{r}\lesssim 5$~kpc, the effect of the galactic
bulge would be important; since we are interested in the star-formation
activities in galactic disks, we ignore the inner regions. For a
comparison, the rotation time of the galaxy at $\tilde{r}=10$~kpc is
shown in figure~\ref{fig:delt}. The minimum time-scale of the evolution
of a galaxy is expected to be the rotation time ($t_{\rm rot}$). For
example, the passage of a galactic arm through a molecular cloud could
stimulate star formation. Thus, if $\Delta t_{\rm 10-5} \lesssim t_{\rm
rot}$, the ram-pressure stripping can be regarded as an instantaneous
phenomenon for the galaxy, which is the case for our model galaxy
(figure~\ref{fig:delt}). Thus, the long-term evolution of the galaxy
does not need to be considered. Moreover, $\Delta t_{\rm 10-5}$ is much
smaller than the crossing time of the galaxy in the cluster. If
turbulent and viscous stripping is considered, the stripping could be
even faster \citep{qui00}. Since the $\rho_{\rm ICM}$ and $v_{\rm rel}$
rapidly decrease as the distance from the cluster center increases, the
ram-pressure becomes less effective in the outer region of the
cluster. Thus, at cluster radii $r> r_{\rm st}$, where $r_{\rm st}$ is
defined by the pressure balance at the galactic radius $\tilde{r}=r_{\rm
gal}$ [equation~(\ref{eq:strip})], it is unlikely that the ram-pressure
directly affects the evolution of the galactic disk. Although galactic
gas at $\tilde{r}>r_{\rm gal}$ might be affected by ram-pressure at
$r>r_{\rm st}$, the effect should be included in `strangulation', which
is considered in Paper~I.

\begin{figure}
  \begin{center}
    \FigureFile(100mm,100mm){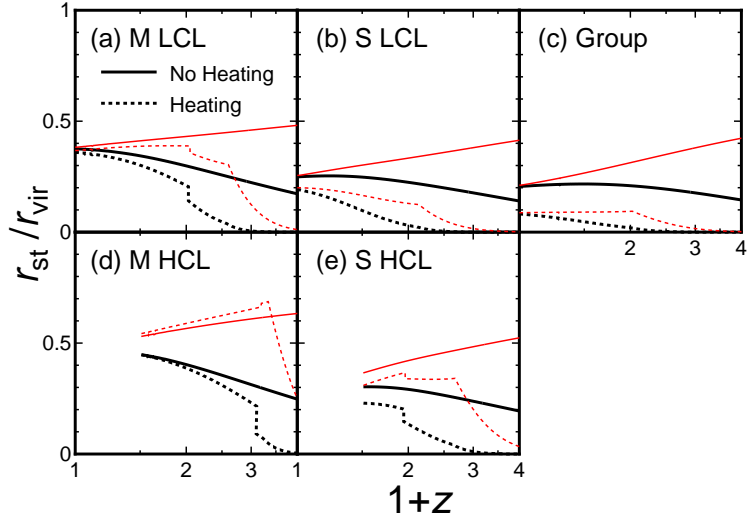}
  \end{center}
  \caption{Evolution of the stripping radii, $r_{\rm st}$, normalized by
the virial radii, $r_{\rm vir}$, for the bigger galaxy in the (a) main
cluster of the LCL, (b) subcluster of the LCL, (c) group, (d) main
cluster of the HCL, and (e) subcluster of the HCL. Solid lines are the
results of the non-heated ICM model and dotted lines are those of the
heated ICM model. Bold and thin lines are those when the evolution of a
galactic disk is considered and those when it is not considered,
respectively, For $r<r_{\rm st}$, ram-pressure stripping is
effective. \label{fig:rst}}
\end{figure}

\begin{figure}
  \begin{center}
    \FigureFile(100mm,100mm){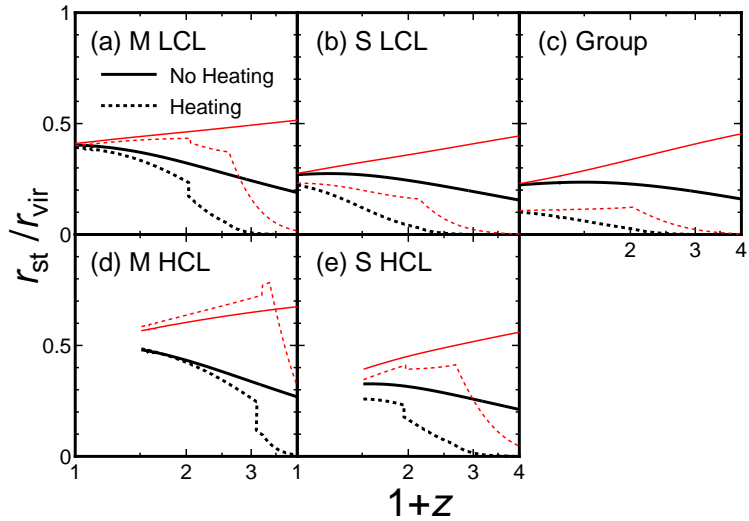}
  \end{center}
  \caption{Same as figure~\ref{fig:rst}, but for the
 smaller galaxy.  \label{fig:rst_s}}
\end{figure}

Since cold gas in a galactic disk is almost instantaneously stripped by
ram-pressure, $r_{\rm st}/r_{\rm vir}$ should be related to the fraction
of galaxies affected by ram-pressure stripping in a cluster. The
evolutions of $r_{\rm st}/r_{\rm vir}$ are shown in
figures~\ref{fig:rst} (the bigger galaxy) and~\ref{fig:rst_s} (the
smaller galaxy).  Compared to the non-heated ICM model, $r_{\rm
st}/r_{\rm vir}$ decreases faster toward higher redshift in the heated
ICM model. This is because the non-gravitational heating reduces the ICM
density and the ram-pressure on galaxies in the inner part of a
cluster. In the heated ICM models, the changes of the slopes correspond
to the transformations of the assumed ICM distributions (see Paper~I).
The differences between figures~\ref{fig:rst} and~\ref{fig:rst_s} are
small, which shows that the differences of the galaxy properties do not
much affect the ram-pressure stripping. Of course, for much smaller
galaxies, $r/r_{\rm st}$ should be much different from those in
figures~\ref{fig:rst} and~\ref{fig:rst_s}. However, those small galaxies
are difficult to be observed in detail at high redshifts.

In figures~\ref{fig:rst} and~\ref{fig:rst_s}, we also present the
evolutions when the disk properties do not evolve, which were shown in
figures~2 and~3 in Paper~I; for $z\lesssim 0.5$, when the disk
properties evolve, $r_{\rm st}/r_{\rm vir}$ is not much different from
that when they do not evolve. For $z\gtrsim 1$, however, the former is
much smaller than the latter. This is because in the former case, the
surface density of the disk is larger at a higher redshift, and the disk
is more robust against ram-pressure stripping
[equation~(\ref{eq:evo_star})]. As discussed in \citet{fuj01a}, the
ram-pressure from the ICM is larger at higher redshifts, because the
average mass density of clusters at higher redshifts is larger than that
of clusters at lower redshifts. Thus, the results show that the effect
of the larger disk surface density dominates that of the larger
ram-pressure at high redshifts.

For the HCL, $r_{\rm st}/r_{\rm vir}$ is larger than that for the LCL
for a given redshift, because the mass of the HCL is larger than that of
the LCL, and the typical galaxy velocity in the HCL is larger than that
in the LCL (Paper~I). On the other hand, for a given mass, ram-pressure
stripping is more effective in higher redshift clusters. The mass of the
main cluster of the HCL at $z\sim 1$ is almost the same as that of the
LCL at $z\sim 0.5$ ($\sim 3\times 10^{14}\: \MO$;
figure~\ref{fig:mass}). However, $r_{\rm st}/r_{\rm vir}$ is larger and
ram-pressure stripping is more effective in the former.

Tormen, Moscardini, and Yoshida (2004) studied the orbital properties of
galaxies in massive clusters at $z<0.8$ by numerical simulations. Since
they showed that the typical peri-centric radius of galaxies for the
first passage is about $0.3\: r_{\rm vir}$, ram-pressure stripping is
effective if $r_{\rm st}/r_{\rm vir}\gtrsim 0.3$. Here, we assume that
the results of Tormen, Moscardini, and Yoshida (2004) can be applied to
our model clusters and the progenitors.  If the disk evolution is
considered and non-gravitational heating is not considered, ram-pressure
stripping is effective at $z\lesssim 1$ for the main cluster of the LCL,
$z\lesssim 2.5$ for the main cluster of the HCL, and $z\lesssim 1$ for
the subclusters of the HCL, while it is not effective for the
subclusters of LCL and the group regardless of $z$
(figures~\ref{fig:rst} and~\ref{fig:rst_s}). Figures~\ref{fig:rst}
and~\ref{fig:rst_s} also show that if both the disk evolution and
non-gravitational heating are considered, ram-pressure stripping is
effective at $z\lesssim 0.5$ for the main cluster of the LCL, and
$z\lesssim 1.5$ for the main cluster of the HCL, while it is not
effective for the subclusters of the LCL and HCL, and the group
regardless of $z$. We note that in Paper~I the tidal force from the main
cluster and the resultant shift of the orbit of a galaxy in the
subcluster are estimated; the estimated peri-centric radius of the
galaxy is $\sim 0.2 r_{\rm vir}$. Thus, the assumed threshold of
ram-pressure stripping ($r_{\rm st}/r_{\rm vir}\lesssim 0.3$) seems to
be reasonable even for the subclusters. For the subclusters of the LCL,
for example, since $0.2\lesssim r_{\rm st}/r_{\rm vir}\lesssim 0.3$ at
$z\lesssim 1$ in the non-heated ICM model , we should rather say that
the ram-pressure stripping is marginally effective at $z\lesssim 1$ in
the non-heated ICM model.

\subsection{Evaporation}
\label{sec:res_eva}

In Paper~I, we mainly investigated the evaporation effect on galaxies
in clusters (or the progenitors) that have not been heated
non-gravitationally. In this study, we also focus on the evaporation in
clusters that have been heated non-gravitationally. Moreover, we study
the saturation effect of thermal conduction on the evaporation
time-scale in detail. Following Paper~I, we assume that $M_{\rm cold,
0}=5\times 10^9\; \MO$ for the bigger galaxy, and $M_{\rm cold,
0}=4\times 10^9\; \MO$ for the smaller galaxy.

\begin{figure}
  \begin{center}
    \FigureFile(100mm,100mm){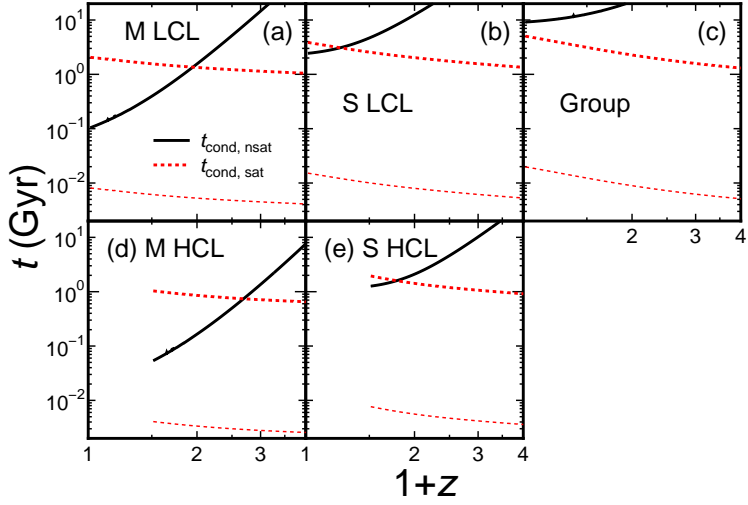}
  \end{center}
  \caption{Evolution of the conduction times for the non-heated ICM
 model for the bigger galaxy in the (a) main cluster of the LCL, (b)
 subcluster of the LCL, (c) group, (d) main cluster of the HCL, and (e)
 subcluster of the HCL.  The solid lines are $t_{\rm cond, nsat}$ at
 $r=0$ (thin lines) and $r=r_{\rm vir}$ (thick lines). The dotted lines
 are $t_{\rm cond, sat}$ at $r=0$ (thin lines) and $r=r_{\rm vir}$
 (thick lines). Note that $t_{\rm cond, nsat}(r=0)=t_{\rm cond,
 nsat}(r_{\rm vir})$ in this figure.  \label{fig:tc2}}
\end{figure}

\begin{figure}
  \begin{center}
    \FigureFile(100mm,100mm){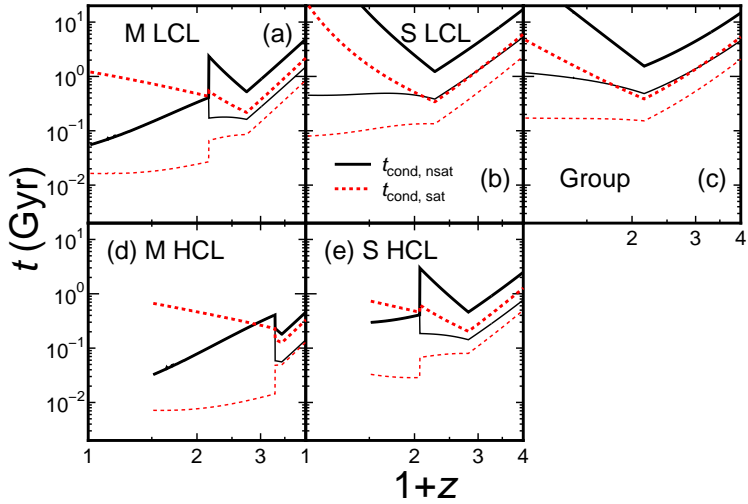}
  \end{center}
  \caption{Same as figure~\ref{fig:tc2}, but for the heated ICM model.
 \label{fig:tc2_h}}
\end{figure}

In figures~\ref{fig:tc2} and~\ref{fig:tc2_h}, we present $t_{\rm cond,
nsat}$ and $t_{\rm cond, sat}$ at $r=0$ and $r_{\rm vir}$ for the bigger
galaxy; figure~\ref{fig:tc2} is for the non-heated ICM model, and
figure~\ref{fig:tc2_h} is for the heated ICM model. Note that since
$t_{\rm cond, nsat}\propto M_{\rm cold}/r_{\rm gal}$ for the unsaturated
case and $t_{\rm cond, sat}\propto M_{\rm cold}/r_{\rm gal}^2$ for the
saturated case [equations~(\ref{eq:cond}), (\ref{eq:Lcond}),
and~(\ref{eq:Lsat})], $t_{\rm cond, nsat}$ ($t_{\rm cond, sat}$) for the
smaller galaxy is 1.6 (3.2) times {\it larger} at a given redshift for
our parameters. Thus, the differences between the bigger galaxy and the
smaller galaxy do not much affect the following results. The actual
evaporation time-scale of a galaxy at a radius $r$ is given by $t_{\rm
cond}(r)\equiv \max[t_{\rm cond, nsat}(r), t_{\rm cond, sat}(r)]$. For
example, at lower redshifts for the main cluster of the LCL, the
saturation cannot be ignored at $r=r_{\rm vir}$ and $t_{\rm cond}=t_{\rm
cond, sat}$.

In the non-heated ICM model, $t_{\rm cond, nsat}$ increases as $z$
increases because $t_{\rm cond, nsat}\propto T_{\rm ICM}^{-5/2}$ and
$T_{\rm ICM}$ decreases rapidly with the mass of the cluster progenitors
[equations~(\ref{eq:cond}) and~(\ref{eq:Lcond})].  In the non-heated ICM
model, $t_{\rm cond, sat}(r=0)$ and $t_{\rm cond, sat}(r=r_{\rm vir})$
decrease slowly as $z$ increases (figure~\ref{fig:tc2}). This is because
their $T_{\rm ICM}$-dependence is small ($\propto T_{\rm ICM}^{-1/2}$)
and the increase of $n_e$ at high redshift dominates
[equations~(\ref{eq:cond}) and~(\ref{eq:Lsat})].

In the heated ICM model, the evolution of the conduction time-scales is
more complicated (figure~\ref{fig:tc2_h}). In general, the conduction
time-scales in the heated ICM model are smaller than those in the
non-heated ICM model because of larger $T_{\rm ICM}$ owing to
non-gravitational heating. The detailed behavior of the conduction
time-scale when thermal conduction is not saturated, $t_{\rm cond,
nsat}$, can be explained as follows. For the main clusters of the LCL
and the HCL and the subcluster of the HCL, $t_{\rm cond, nsat}$
increases as $z$ increases. This is because $T_{\rm ICM}$ decreases as
is the case of the non-heated ICM model. However, the increasing rate is
smaller because $T_{\rm ICM}$ decreases more slowly owing to
non-gravitational heating. For higher redshifts, the viral temperature
becomes smaller and satisfies the relation $T_{\rm vir}<T_0$, where
$T_0$ (=0.8~keV) is the critical temperature under which the ICM
distribution follows an adiabatic accretion model (subsection
\ref{sec:model_ram}). For example, $T_{\rm vir}$ equals $T_0=0.8$~keV at
$z=1.2$ for the main cluster of the LCL. Since the cluster is not
isothermal in the adiabatic accretion model [equation~(\ref{eq:T_ad})],
$t_{\rm cond, nsat}$ bifurcates there. For the subcluster of the LCL and
the group, the density and temperature profiles follow those predicated
by the adiabatic accretion model for $z\ge 0$. In these redshift ranges,
$t_{\rm cond, nsat}(r=r_{\rm vir})>t_{\rm cond, nsat}(r=0)$ because
$T_{\rm ICM}(r=r_{\rm vir})<T_{\rm ICM}(r=0)$. In the middle-redshift
range, $1.2<z<1.7$ for the main cluster of the LCL for example, the ICM
fraction of a cluster is the same as that at lower redshift although the
cluster is more compact at the higher redshifts. Therefore, $T_{\rm
ICM}$ increases via adiabatic compression and $t_{\rm cond,
nsat}(r=r_{\rm vir})$ decreases as $z$ increases. At higher redshifts,
$z>1.7$ for the main cluster of the LCL for example, $t_{\rm cond,
nsat}$ increases as $z$ increases, because the ICM fraction and $T_{\rm
ICM}$ decrease.

The behavior of the saturated conduction time-scale, $t_{\rm cond,
sat}$, at lower redshifts ($z\lesssim 1.2$ for the main cluster of the
LCL, for example) can be explained as follows. Since the average dark
matter and ICM densities increase toward higher redshifts, the increase
of $t_{\rm cond, sat}$ through the decrease of $T_{\rm ICM}$ is
dominated by the decrease of $t_{\rm cond, sat}$ through the increase of
$n_e$ [equations~(\ref{eq:cond}) and~(\ref{eq:Lsat})] at $r=r_{\rm
vir}$. However, in our ICM model, for a given gas mass fraction, the
non-gravitational heating decreases the ICM density in the central
region of a cluster.  Thus, the decrease of $t_{\rm cond, sat}$ is not
significant at $r=0$. At higher redshifts, once the density and
temperature profiles follow those predicted by the adiabatic accretion
model, the behaviors of $t_{\rm cond, sat}$ can be attributed to the
mechanisms that are the same as the case of $t_{\rm cond, nsat}$.

\begin{figure}
  \begin{center}
    \FigureFile(100mm,100mm){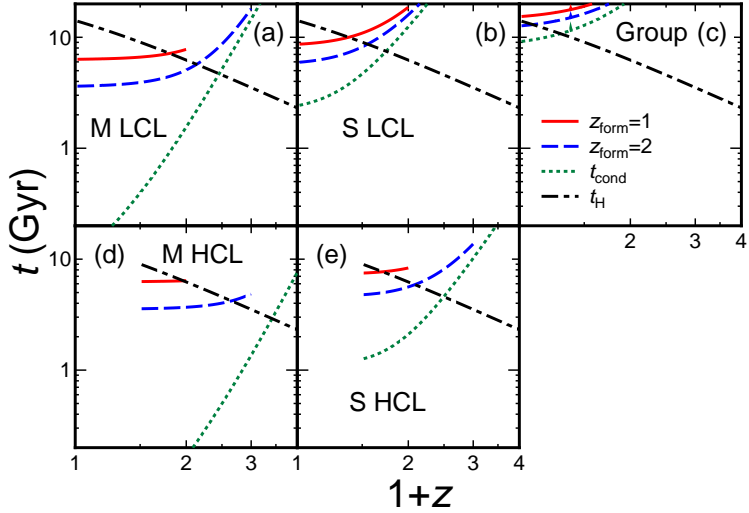}
  \end{center}
  \caption{Time-scales of thermal conduction, $t_{\rm cond}(r=0)$ for
 the non-heated ICM model for the bigger galaxy in the (a) main cluster
 of the LCL, (b) subcluster of the LCL, (c) group, (d) main cluster of
 the HCL, and (e) subcluster of the HCL. The solid and dashed lines show
 $t_{\rm cond}(r=0)'=t_{\rm cond}(r=0)+t_{\rm form}$ when $z_{\rm
 form}=1$ and~2, respectively. The dotted and dot-dashed lines show
 $t_{\rm cond}(r=0)$ and the Hubble time $t_{\rm H}$, respectively.  For
 $t_{\rm cond}'\lesssim t_{\rm H}$, thermal conduction is
 effective. \label{fig:tcond}}
\end{figure}

In figure~\ref{fig:tcond}, we show the evolution of $t_{\rm cond}=
\max(t_{\rm cond, nsat}, t_{\rm cond, sat})$ at $r=0$ for the bigger
galaxy in the non-heated ICM model. This figure corresponds to figure~5
in Paper~I in which the evolution of the disk is not taken into
account. As shown in figure~\ref{fig:tc2}, $t_{\rm cond, nsat}\gg t_{\rm
cond, sat}(r=0)$ and therefore $t_{\rm cond}(r=0)=t_{\rm cond,
nsat}$. The conduction time-scale can be represented by $t_{\rm
cond}\propto M_{\rm cold}/r_{\rm gal}$ [equations~(\ref{eq:cond})
and~(\ref{eq:Lcond})]. Since both $M_{\rm cold}$ and $r_{\rm gal}$ are
proportional to $H(z)^{-1}$ [equations~(\ref{eq:evo_rad})
and~(\ref{eq:evo_Mc})], $t_{\rm cond}$ is not influenced by the effect of
the disk evolution. Thus, the upper figures are the same as figures
5a--c in Paper~I. In figure~\ref{fig:tcond}, we also present the Hubble
time, $t_{\rm H}$. If $t_{\rm cond}\ll t_{\rm H}$, the cold gas in a
galaxy is evaporated soon after the galaxy forms in the cluster or its
progenitors. We also show $t_{\rm cond}'(r=0)=t_{\rm cond}(r=0)+t_{\rm
form}$, where $t_{\rm form}$ is the Hubble time when the galaxy forms at
$z=z_{\rm form}$. If $t_{\rm cond}'<t_{\rm H}$ at a given redshift, the
cold gas in the galaxy has been evaporated by the redshift. Note that
$t_{\rm cond}$ corresponds to $t_{\rm cond}'$ when $z_{\rm
form}=\infty$.

Figure~\ref{fig:tcond} shows that in the main cluster of the LCL (HCL)
observed at $z\sim 0.5$ (1), cold gas has been evaporated from the
galaxies near the cluster center. On the other hand, in the subclusters
of the LCL (HCL) observed at $z\sim 0.5$ (1), cold gas is evaporating if
the galaxies formed at $z\sim 1$--2. The time-scales of the evaporation,
$t_{\rm cond}$, is relatively long ($\gtrsim 2$~Gyr). In the group, cold
gas is evaporating at $z\sim 0$.

\begin{figure}
  \begin{center}
    \FigureFile(100mm,100mm){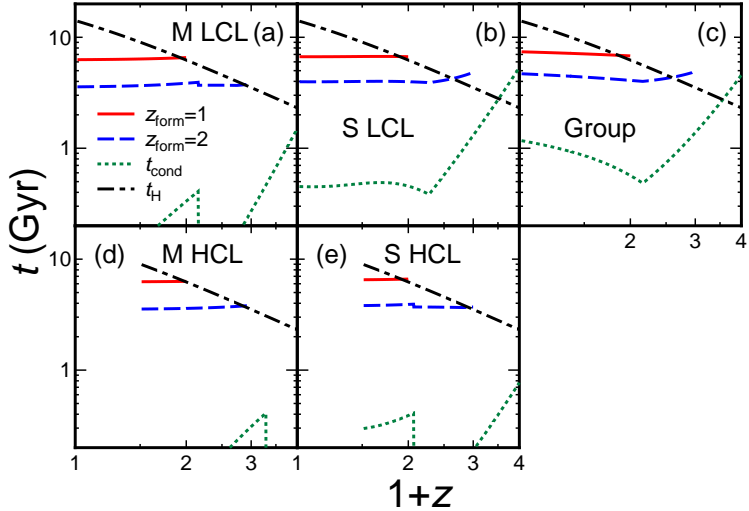}
  \end{center}
  \caption{Same for figure~\ref{fig:tcond}, but for
 the heated ICM model. \label{fig:tcond_h}}
\end{figure}

In figure~\ref{fig:tcond_h}, we present the evolutions of $t_{\rm
cond}(r=0)$ and $t_{\rm cond}'(r=0)$ for the bigger galaxy in the heated
ICM model. Since $t_{\rm cond}$ in the heated ICM model is generally
smaller than that in the non-heated ICM owing to the larger $T_{\rm
ICM}$, $t_{\rm cond}'$ becomes smaller than $t_{\rm H}$ soon after
galaxy formation.

A galaxy that first enters a cluster from the outside does not stay at
$r=r_{\rm vir}$ and the orbit is not exactly radial. If we assume that
the typical peri-centric radius of galaxies is about $0.3\: r_{\rm
vir}$, ram-pressure stripping is effective for most galaxies if $r_{\rm
st}/r_{\rm vir}\gtrsim 0.3$ (see subsection~\ref{sec:ramp}). On the
other hand, if $r_{\rm st}/r_{\rm vir}\lesssim 0.3$, the galaxies would
be affected only by evaporation while they are orbiting.  In this case,
the actual time-scale of evaporation, $t_{\rm evap}$, is given by
$t_{\rm cond}(r=0)<t_{\rm evap}<t_{\rm cond}(r=r_{\rm vir})$.
Therefore, in the subclusters of the LCL observed at $z\sim 0.5$, the
bigger galaxy infalling from the outside loses its cold gas with a
time-scale of $0.5<t_{\rm evap}<10$~Gyr in the heated ICM model
(figure~\ref{fig:tc2_h}). In the subclusters of the HCL observed at
$z\sim 1$, a galaxy loses its cold gas with the time-scale of $t_{\rm
evap}\sim 0.5$~Gyr in the heated ICM model (figure~\ref{fig:tc2_h}).

\subsection{Comparison between Ram-pressure Stripping and Evaporation}

Although both ram-pressure stripping and evaporation suppress
star-formation activities in galaxies, they affect the star-formation
activities differently. For position, while the ram-pressure stripping
is effective only in the central regions of clusters ($r\lesssim 0.5
r_{\rm vir}$; figures~\ref{fig:rst} and~\ref{fig:rst_s}), the
evaporation is often effective even at $r\sim r_{\rm vir}$. For
time-scale, while the ram-pressure stripping checks star-formation
activities in a very short time ($\sim 10^8$~yr; figure~\ref{fig:delt},
see Fujita, Nagashima 1999), the evaporation generally affects the
star-formation activities more slowly ($\sim 10^9$~yr;
figures~\ref{fig:tc2} and~\ref{fig:tc2_h}). The difference in the
decline rate of the star-formation activities could be discriminated by
the spectra of the galaxies. These facts would be useful to
observationally find whether ram-pressure stripping or evaporation
dominates in clusters. In the next section, we investigate several
specific cases.

\section{Discussion}
\label{sec:disc}

In Paper~I, the star-formation activities of galaxies in main clusters
and the subclusters in the vicinity of the main clusters observed at
$z\sim 0.5$ are discussed. Since the consideration of the disk evolution
does not much change $r_{\rm st}/r_{\rm vir}$ for $z\lesssim 0.5$ for
the LCL (figures~\ref{fig:rst} and~\ref{fig:rst_s}), the conclusions
about the effects of ram-pressure stripping on the galaxies do not
change either.  In Paper~I, we discussed that the rapid decline of the
star-formation rates of galaxies in the main cluster of the LCL is not
consistent with observations of the CNOC sample of very luminous X-ray
clusters \citep{bal99b,kod01a} and this suggests that the star-formation
rates have decreased before the galaxies enter the main clusters
\citep{got03b,got03c}. In Paper~I, we argued that the `pre-processing'
occurred in the subclusters \citep{zab98,has00,bal00}, which is
consistent with the observations showing that red galaxies have a clumpy
distribution around a main-cluster \citep{kod01c}. In the subclusters of
the LCL at $z\sim 0.5$, ram-pressure stripping is marginally effective
($0.2\lesssim r_{\rm st}/r_{\rm vir}\lesssim 0.3$) in the non-heated
ICM, and it is ineffective in the heated ICM model
(figures~\ref{fig:rst} and~\ref{fig:rst_s}). If ram-pressure stripping
is the mechanism of the pre-processing, most of the cold gas in galaxies
may be removed and star-formation activity of the galaxies may
completely die out before the galaxies enter the main cluster. This may
be inconsistent with the existence of many blue galaxies in main
clusters at $z\lesssim 0.5$ \citep{kod01a}; the observations suggest a
slower decline of the star-formation rates.  Thus, the heated ICM model
is preferable because ram-pressure stripping can be ignored.  In the
heated ICM model, evaporation can be candidates of the mechanism of the
pre-processing. The time-scale of the evaporation is $0.5\lesssim t_{\rm
evap}\lesssim 10$~Gyr. Since the time-scale is relatively large, the
evaporation can be an alternative of strangulation, which is expected to
gradually suppress the star-formation activities of galaxies
\citep{lar80,kod01c,bek02}, but is highly speculative \citep{ben00}.

For the HCL observed at $z\sim 1$, ram-pressure stripping is effective
in the main cluster regardless of non-gravitational heating. As is the
case of the LCL at $z\sim 0.5$, the ICM of the subclusters must have
been heated to avoid ram-pressure stripping in the subclusters.

At higher redshifts ($z\gtrsim 1$), the effect of the disk evolution is
more significant (figures~\ref{fig:rst} and~\ref{fig:rst_s}), although
it is not well-known that disk galaxies with the rotation velocities
that we studied exist. At these redshifts, most of the subclusters have
not been absorbed in the main cluster. As long as the ICM has not been
heated non-gravitationally, the efficiency of ram-pressure stripping
decreases only slowly as redshift increases. Thus, the products of
ram-pressure stripping, such as galaxies that have spectra reflecting
rapidly declined star-formation rates, may be observed at these
redshifts although the number fraction is not as much as that at lower
redshifts. In the non-heated ICM model, figure~\ref{fig:tcond} shows
that the evaporation is ineffective ($t_{\rm cond}\gtrsim t_{\rm H}$)
for less massive systems such as the subclusters of the LCL (at $z>0.7$)
and the HCL (at $z>1.5$). Thus, if the evaporation is the mechanism of
the pre-processing, the star-formation rates of galaxies hardly decline
in the vicinity of the clusters at $z\gtrsim 1$--2 except for galaxies
in which star-formation activities decrease rapidly by the ram-pressure
stripping. On the other hand, if the ICM has been heated, ram-pressure
stripping does not occur in the high-redshift range
(figures~\ref{fig:rst} and~\ref{fig:rst_s}).  The time-scale of the
evaporation is given by $t_{\rm cond}(r=0)\lesssim t_{\rm evap}\lesssim
t_{\rm cond}(r_{\rm vir})$ and is relatively large; for example,
$1\lesssim t_{\rm evap}\lesssim 4$~Gyr at $z\sim 2$ for the subcluster
of the LCL (figure~\ref{fig:tc2_h}). Thus, the star-formation rates of
galaxies should decrease slowly with this time-scale.  Thus, the heated
and non-heated ICM models may be discriminated by future observations of
the star-formation history of galaxies in the regions surrounding main
clusters. However, quantitative predictions based on semi-analytic
models or numerical simulations would be required for actual
discrimination. We leave this for future study.

\section{Conclusions}
\label{sec:conc}

We have investigated ram-pressure stripping and evaporation of disk
galaxies in and around a cluster using analytical models based on a
hierarchical clustering scenario. We considered the redshift evolution
of the size and surface density of the disk. We showed that the
evolution does not much affect the efficiency of ram-pressure stripping
of galaxies for $z\lesssim 0.5$, but affects for $z\gtrsim 1$. We also
considered the saturation of thermal conduction in detail, and found
that it cannot be ignored at larger radii of a cluster, which makes the
time-scale of the evaporation larger. Thus, the evaporation has the same
effect as strangulation (stripping of warm gas in galactic halos) to
suppress the star-formation activities of galaxies
gradually. Observations of galaxies in the vicinity of clusters at
$z\sim 1$ are useful to investigate whether non-gravitational heating
has occurred or not by $z\sim 1$.

\vspace{5mm}

We are grateful to M. Nagashima and M. Enoki for useful comments.
Y.~F. was supported in part by a Grant-in-Aid from the Ministry of
Education, Culture, Sports, Science and Technology (14740175).


\end{document}